\documentclass{ws-procs975x65}            
\begin{document}
\title{Old Wine in a New Bottle: \\
Technidilaton as the 125 GeV Higgs\\ 
-- Dedicated to the late Professor Yoichiro Nambu\footnote{deceased on July 5, 2015}---
}\footnote{to appear in Proceedings of the Sakata Memorial KMI Workshop on ``Origin of Mass and Strong Coupling Gauge Theories''
(SCGT15), March 3-6, 2015, KMI, Nagoya University, Nagoya, Japan}

\author{Koichi Yamawaki
}

\address{Kobayashi-Maskawa Institute for the Origin of Particles and the Universe (KMI),\\ Nagoya University, 
Nagoya, 464-8602, Japan.\\
}



\begin{abstract}
-- One theme that emerged from these observations is that some concepts have a long life. They may not always be right, but they can undergo reincarnations and become relevant later (``old wine in a new bottle'', a remark we have heard).---Y. Nambu,
Concluding Remarks at SCGT 88.---\\

The first Nagoya SCGT workshop  back in 1988 (SCGT 88) was motivated by 
the walking technicolor and technidilaton.
Now at 
SCGT15 
I returned to the ``old wine'' 
in ``a new bottle'',  the 
recently discovered 125 Higgs boson as the technidilaton. We show that the Standard Model (SM) Higgs Lagrangian is identical to
the nonlinear realization of both the scale and chiral symmetries (``scale-invariant nonlinear sigma model''), and is further gauge equivalent to the ``scale-invariant Hidden Local Symmetry (HLS) model'' having possible new vector bosons as the HLS gauge bosons with scale-invariant mass:
SM Higgs is nothing but a (pseudo) dilaton.  The effective theory of the walking technicolor has precisely the same type of the scale-invariant nonlinear sigma model,  thus further having the scale-invariant HLS gauge bosons (technirho's, etc.).  The technidilaton mass $M_\phi$ comes from the trace anomaly, which yields 
$M_\phi^2 F_\phi^2 \simeq  [\frac{8}{N_F}\frac{4}{N_C}]\cdot (2.5)^2  \cdot v^4$ via PCDC, in  the underlying walking $SU(N_C)$ gauge theory with $N_F$ massless flavors,
where $F_\phi$ is the the decay constant and $v=246$ GeV.
 This implies $F_\phi\simeq 5 \,v$ for  
$M_\phi \simeq
125\, {\rm GeV}\simeq \frac{1}{2} v$ in the 
one-family walking technicolor model ($N_C=4, N_F=8$), 
in good agreement with the current LHC Higgs data. 
In the anti-Veneziano limit, $N_C \rightarrow \infty$, with $N_C \alpha={\rm fixed}$ and
$N_F/N_C={\rm fixed} \, (\gg 1)$, 
we have a result: 
 $M_\phi^2/v^2\sim M_\phi^2/F_\phi^2   \sim 1/(N_F N_C) \rightarrow 0 $.  
 Then the  technidilaton is a  naturally light composite Higgs out of the strongly coupled conformal dynamics, with its couplings 
even weaker than the SM Higgs. 
 Related holographic and lattice results are also discussed. In particular, such a light flavor-singlet scalar does exists in the lattice
 simulations in the walking regime.
\end{abstract}


\bodymatter

\section{Introduction}
\label{aba:sec1}

 This talk is an impromptu substitute filling the hole of the cancellation of the talk to be given by Volodya Miransky, my old friend, so
 this is a kind of personal sentiments style rather than a scientific presentation. Sorry for that in advance.
 
History repeats itself:  At the first SCGT workshop in 1988 motivated by our work \cite{Yamawaki:1985zg,Bando:1986bg} proposing the walking technicolor based on his paper \cite{Miransky:1984ef}, 
the Volodya's talk was also cancelled, although he was a frequent repeater to many (out of ten)  SCGT
workshops  during the long period more than a quarter century 1988 - 2015. 

Volodya discovered  \cite{Miransky:1984ef}  so-called Miransky scaling:
 \begin{equation}
 m_F \simeq 4  \Lambda\cdot  \exp \left(-\frac{\pi}{\sqrt{\frac{\alpha}{\alpha_{\rm cr}}-1}}\right) \, \ll \Lambda \quad \left(\frac{\alpha}{\alpha_{\rm cr}}-1\ll 1\right)  
 \label{Miranskys} 
 \end{equation}
 in the scale-invariant gauge model (ladder model),  an essential singularity scaling analogous to the Berezinsky-Kosterlitz-Thouless phase transition (what we called ``conformal phase transition'' \cite{Miransky:1996pd}),  where $m_F (\ne 0)$ is the spontaneous chiral symmetry breaking (S$\chi$SB) 
solution of the ladder Schwinger-Dyson (SD) gap equation for the fermion and  $\Lambda$ is the ultraviolet cutoff to regularize the theory to act as an intrinsic scale $\Lambda_{\rm TC}$ of the walking technicolor, similarly to the $\Lambda_{\rm QCD}$ in QCD,  responsible for the trace anomaly $\theta_\mu^\mu \propto \Lambda^4$ as the explicit breaking of the scale symmetry manifesting itself as  the running of the coupling in the ultraviolet region far bigger than the S$\chi$SB scale $m_F$. 

The S$\chi$SB solution  exists only in the {\it strong coupling} phase $\alpha>\alpha_{\rm cr}\ne 0$, 
where the {\it non-zero critical coupling $\alpha_{\rm cr}\ne 0$}
discovered by Maskawa-Nakajima \cite{Maskawa:1974vs} is the characterization of the {\it strong coupling gauge theories (SCGT)}. It is a gauge analogue of the {\it non-zero critical coupling, $G^{(\rm NJL)} >G_{\rm cr}\ne 0$, of the Nambu-Jona-Lasinio (NJL) model} \cite{Nambu:1961tp}, which, despite the resemblance,  should actually be {\it distinguished from the BCS dynamics having zero critical coupling} $G_{\rm cr}^{(\rm BCS)} =0$ (weak coupling theory in
broken phase even for infinitesimal coupling).\footnote{ The characteristic weak coupling of the BCS theory is   due to the existence of the Fermi surface where the fermions are effectively in the 2-dimensional brane instead of 4 dimensional free space bulk, whereas the Cooper pairs and Nambu-Goldstone (NG) modes are bosons and hence live in the bulk to trigger the Higgs mechanism (Meissner effect), escaping the Mermin-Wagner-Coleman theorem in genuine 2 dimensional theories.
}

Volodya further invented  nonperturbative renormalization \cite{Miransky:1984ef}   taking $\Lambda \rightarrow \infty$ as $m_F={\rm fixed}$ for the Miransky scaling Eq.(\ref{Miranskys}), which yields a nonperturbative beta function: \cite{Leung:1985sn}
\begin{eqnarray}
\beta^{(NP)}(\alpha) &=&\Lambda \frac{\partial \alpha(\Lambda)}{\partial \Lambda}\Bigg|_{m_F} = - \frac{2\pi^2\alpha_{\rm cr}}{\ln^3 (\frac{4\Lambda}{m_F})} 
=  - \frac{2\alpha_{\rm cr}}{\pi} \left(\frac{\alpha}{\alpha_{\rm cr}}-1\right)^{\frac{3}{2}} 
\label{NPbeta},\\
\alpha(\mu) &=&\alpha_{\rm cr} 
\left[1+ \frac{\pi^2}{\ln^2(\frac{\mu}{m_F})}
\right] \,,\quad  (\alpha> \alpha_{\rm cr}\ne 0)\label{NPrun}
\end{eqnarray}
Thus the coupling actually starts running nonperturbatively once the $m_F\ne 0$ is generated,  even
if the initial ladder coupling is scale-invariant (nonrunning).  Besides the explicit breaking by the intrinsic scale $\Lambda$, the scale symmetry is now {\it explicitly broken} also by $m_F$ which was generated by the {\it spontaneous breaking} of the scale symmetry, with the {\it new trace anomaly
(nonperturbative trace anomaly)} $|{\theta_\mu^\mu}^{(NP)}| ={\cal O}(m_F^4) (\ll \Lambda^4)$. 
From Eq.(\ref{NPbeta}) we can see that the  $\alpha_{\rm cr}\ne 0$ is regarded as  a {\it nontrivial ultraviolet fixed point}, although the input ladder coupling $\alpha(\mu^2)=\alpha$ in
the infrared region $\mu^2<\Lambda^2=\Lambda_{\rm TC}^2$
is regarded as the {\it nontrivial infrared fixed point}. 
 This opened a new phase of the SCGT for the composite model and turned out to be the basics of the whole walking dynamics. 

We applied this Volodya's result to the technicolor theory, and proposed what we called ``Scale-invariant Technicolor'' (now called Walking Technicolor)  \cite{Yamawaki:1985zg,Bando:1986bg}, 
 where  we found a large anomalous dimension 
\begin{equation}
\gamma_m=1\,\quad\quad  (\alpha> \alpha_{\rm cr})
\label{gamma}
\end{equation}
in the broken phase\footnote{The anomalous dimension in the unbroken phase ($\alpha<\alpha_{\rm cr}$) was known \cite{Leung:1985sn}
  to be $\gamma_m=1-\sqrt{1-\alpha/\alpha_{\rm cr}}$, which is irrelevant  to the dynamical mass $m_F$ ($\ne 0$ only for $\alpha>\alpha_{\rm cr}$) and hence to the nonperturbative running of the coupling in Eq.(\ref{NPbeta}).
},  as a solution of the Flavor-Changing-Neutral Currents (FCNC) problem of the original Technicolor \cite{TC},\footnote{The FCNC solution by the large anomalous dimension was proposed without concrete dynamical model nor concrete value of the anomalous dimension at a hypothetical nontrivial fixed point in the asymptotically non-free theory.\cite{Holdom:1981rm}  Subsequently to our paper, a similar FCNC solution based on the ladder-type SD equation was discussed \cite{Akiba:1985rr}, without notion of the anomalous dimension nor scale symmetry, and hence without  technidilaton.   
} and predicted a technidilaton, a pseudo Nambu-Goldstone boson of the approximate scale symmetry, which is spontaneously broken at the same time as the S$\chi$SB due to the fermionic condensate responsible for the electroweak symmetry breaking. {\it The technidilaton was predicted as a flavor-singlet 
technifermion bound state} (not a glueball-type bound state), behaving similarly to the standard model (SM) Higgs itself. 

Below I will explain \cite{Matsuzaki:2015sya} how the
technidilaton is a {\it naturally light and weakly coupled composite Higgs} out of {\it strongly coupled} underlying conformal gauge theory, the walking technicolor, in the light of the anti-Veneziano limit $N_C, N_F \rightarrow \infty$
with $N_F/N_C=$fixed $\gg 1$ for $SU(N_C)$ gauge theory with $N_F$ massless flavors.  The technidilaton particularly for the one-family walking technicolor with $N_F=8$ and $N_C=4$ is nicely fit to the current 125 GeV Higgs data at LHC. \cite{Matsuzaki:2012mk,Matsuzaki:2012xx,Matsuzaki:2015sya}

The present SCGT15 workshop is entitled ``Sakata Memorial $\cdots$''.  Composite model is indeed Nagoya University tradition trace back to the late Professor Shoichi Sakata, who invented the Sakata model \cite{Sakata:1956hs},
a composite hadron model, in 1955 (published in 1956), which turned out to be the forerunner of the quark model in 1964 \cite{GellMann:1964nj}. In the context of the extended Sakata model with the four constituents corresponding to the four lepton flavors, the famous Maki-Nakagawa-Sakata (MNS) neutrino mixing was proposed in 1962 \cite{Maki:1962mu}. This motivated M. Kobayashi and T. Maskawa, both disciples of Sakata, to believe in four quarks, with quark flavor mixing counter  to the MNS lepton flavor mixing, which was the initial motivation of the paper of Kobayashi-Maskawa \cite{Kobayashi:1973fv}, stepping eventually further to the six quarks in 1972 (published in 1973), well before the $J/\psi$ discovery in 1974. 

The present SCGT15 workshop is also entitled ``Origin of Mass $\cdots$''. The origin of mass of all the SM particles is the Higgs VEV $v=\sqrt{\frac{-\mu_0^2}{\lambda}}=246\,{\rm GeV}$ or the Higgs mass $M_\phi^2 =2\lambda v^2=-2\mu_0^2$ read from  the Lagrangian:
\begin{equation}
 {\cal L}_{\rm Higgs}= |\partial_\mu h|^2 -\mu_0^2 |h|^2 -\lambda|h|^4
\,. 
\label{Higgs}
\end{equation}
Then the origin of mass 
is attributed to  the mysterious input mass parameter of the {\it tachyon} with the mass $\mu_0$ such that $\mu_0^2<0$ {\it as a free parameter}.
But why tachyon? How is the tachyon mass determined? SM cannot answer to these questions. 

Here we should
 recall that 
{\it the spontaneous symmetry breaking was born as a dynamical symmetry breaking, thanks to Professor Y. Nambu} \cite{Nambu:1961tp}, where the tachyon is in fact generated as a dynamical consequence of the strong dynamics, but not an ad hoc input.
Before we discuss the dynamical origin of mass a la Nambu, we first show \cite{Fukano:2015zua}  that {\it  the SM Higgs Lagrangian itself possesses ``hidden'' symmetries (scale symmetry and gauge symmetry, both spontaneously broken, i.e., nonlinearly realized)},  in addition to the well-known symmetry (chiral symmetry, also spontaneously broken)
to be gauged by the electroweak symmetry.

\section{SM Higgs as a Dilaton: Hidden Scale Symmetry and Hidden Local Symmetry in the SM Higgs
Lagrangian \cite{Fukano:2015zua}}

Here we show \cite{Fukano:2015zua} that   the SM Higgs 
Lagrangian Eq.(\ref{Higgs}) in the form of the {\it linear sigma model} 
 is rewritten into precisely the form equivalent to the {\it scale-invariant} version of the 
chiral $SU(2)_L\times SU(2)_R$ {\it nonlinear sigma model}, as far as it is in the broken phase, with {\it both the  chiral and scale symmetries spontaneously broken} due to the same Higgs VEV
$v\ne 0$, and thus are {\it both nonlinearly realized}. The SM Higgs Lagrangian is further shown to be {\it gauge equivalent} to the {\it scale-invariant} 
version \cite{Kurachi:2014qma} of the Hidden Local Symmetry (HLS) Lagrangian \cite{Bando:1984ej,Bando:1987br,Harada:2003jx}, which contains {\it possible new vector bosons}, analogue of the $\rho$ mesons,  as the gauge bosons of the (spontaneously broken) HLS {\it hidden behind the SM Higgs Lagrangian}.

Let us rewrite Eq.(\ref{Higgs})  
as
\begin{eqnarray}
{\cal L}_{\rm Higgs}&=& \frac{1}{2} \left[
\left(\partial_\mu {\hat \sigma}\right)^2 +\left(\partial_\mu {\hat \pi_a}\right)^2
\right]
-\frac{1}{2}
\mu_0^2  \left[ 
{\hat \sigma}^2+{\hat \pi_a}^2 
\right]-\frac{\lambda}{4} \left[ 
{\hat \sigma}^2+{\hat \pi_a}^2 
\right]^2  \nonumber \\
&=& \frac{1}{2} {\rm tr} \left( \partial_\mu M\partial^\mu M^\dagger \right)
 - \left[\frac{\mu_0^2}{2} {\rm tr}\left(M M^\dagger\right)+\frac{\lambda}{4}  \left({\rm tr}\left(M M^\dagger\right)\right)^2\right] \,. 
 \label{Lag:sigma}
 \end{eqnarray}
where we have noted
\begin{equation}
h=\left(\begin{array}{c}
\phi^+\\
\phi^0\end{array}  \right)=\frac{1}{\sqrt{2}} \left(\begin{array}{c}
i{\hat \pi}_1+{\hat \pi}_2\\
\hat{\sigma}-i {\hat \pi}_3\end{array}\right)\,,
\end{equation} 
and $2\times 2$ matrix $M$ reads
\begin{equation}
M=(i \tau_2 h^*, h) = \frac{1}{\sqrt{2}}\left({\hat \sigma}\cdot  1_{2\times 2} +2i {\hat \pi}\right)\, \quad \left({\hat \pi} \equiv {\hat \pi}_a \frac{\tau_a}{2}\right)\,,
\end{equation}
which transforms under $G=SU(2)_L\times SU(2)_R$ as:
\begin{equation}
M \rightarrow g_L \, M\, g_R^\dagger \,,\quad \left(g_{R,L} \in SU(2)_{R,L}\right)\,.
\end{equation}
Note first that any complex matrix $M$ can be decomposed into the Hermitian (always diagnonalizable) matrix $H$  and unitary matrix $U$ as $M=HU$ ( ``polar decomposition'' ): 
 \begin{equation}
 M = H\cdot U\,, \quad H=\frac{1}{\sqrt{2}} \left(\begin{array}{cc}
 \sigma & 0\\
 0  &\sigma
 \end{array}\right)
 \,, \quad U= \exp\left(\frac{2i \pi}{F_\pi}\right) \quad \left(F_\pi=v=\langle  \sigma \rangle\right) \,.
 \label{Polar}
 \end{equation}
 The chiral transformation of $M$ is inherited by $U$, while $H$ is a chiral singlet such that:
 \begin{equation}
 U \rightarrow g_L \, U\, g_R^\dagger\,,\quad H \rightarrow H\,,
 \end{equation}
 and $U \, U^\dagger=1$ implies $\langle U\rangle =\langle  \exp\left(\frac{2i \pi}{F_\pi}\right)\rangle=1 \ne 0$, namely the spontaneous breaking of the chiral symmetry is taken granted in the polar decomposition.
 Note that the radial mode $\sigma$ is a chiral-singlet in contrast to $\hat \sigma$ which is chiral non-singlet.  
 
 We further parametrize $\sigma$  as 
 \begin{equation} 
 \sigma =v \cdot \chi\,,\quad \chi=\exp\left(\frac{\phi}{F_\phi}
 \right)
 \,,
 \label{NLscale}
 \end{equation}
 where $F_\phi=v$ is the decay constant of the dilaton $\phi$ as the Higgs. 
 The scale (dilatation) transformations for these fields are 
 \begin{equation}
 \delta_D \sigma =(1 +x^\mu \partial_\mu) \sigma \,, \qquad  
\delta_D \chi=(1+x^\mu \partial_\mu) \chi\,, \qquad 
\delta_D \phi= F_\phi+x^\mu \partial_\mu\phi\,. 
 \end{equation}
Note that $\langle  \sigma\rangle= v \langle \chi \rangle = v\ne 0$ breaks spontaneously the scale symmetry, but not the chiral symmetry, since 
$ \sigma$ ($\chi$ as well) is a chiral singlet.
This is a nonlinear realization of the scale symmetry: 
the $\phi$ is a dilaton, NG boson of the spontaneously broken scale symmetry. Although $\chi$ is a dimensionless field,
it transforms as that of dimension 1, while $\phi$ having dimension 1 transforms as the dimension 0, instead.
 
 Then the SM Higgs Lagrangian reads:\cite{Fukano:2015zua}
 \begin{eqnarray}
  {\cal L}_{\rm Higgs}
 &=&\left[ \frac{F_\phi^2}{2} \left(\partial_\mu \chi \right)^2+ \frac{F_\pi^2}{4}{\chi}^2\cdot {\rm tr} \left(\partial_\mu U \partial^\mu U^\dagger\right)\right]
  - V(\phi)\nonumber\\
 &=&\chi^2 \cdot \left[ \frac{1}{2} \left(\partial_\mu \phi\right)^2  +\frac{F_\pi^2}{4}{\rm tr} \left(\partial_\mu U \partial^\mu U^\dagger\right)\right] -V(\phi)\nonumber \,,\\
 V(\phi)&=& \frac{\lambda}{4} v^4 \left[\left(\chi^2 -1\right)^2-1\right]  =\frac{M_\phi^2 F_\phi^2}{8} \left[\left(\chi^2 -1\right)^2-1\right] \,,
 \label{SNLSM}
 \end{eqnarray}
which  is nothing but {\it the scale-invariant nonlinear sigma model}~\cite{Matsuzaki:2012mk,Matsuzaki:2015sya}, with $F_\phi=F_\pi=v$,   plus the explicit scale-symmetry breaking potential $V(\phi)$ such that $\delta_D V(\phi) = \lambda v^4\chi^2=-\theta_\mu^\mu$  whose scale dimension $d_{\theta}=2$ (originally the tachyon mass term)  instead of 4: namely, the scale symmetry is broken only by the dimension 2 operator.\footnote{Note that {\it mass of all the SM particles except the Higgs is scale-invariant}. By the electro-weak gauging as usual; $\partial_\mu U\Rightarrow {\cal D}_\mu U= \partial_\mu U -i g_2 W_\mu U +i g_Y U B_\mu$
 in Eq.(\ref{SNLSM}), we see that the mass of $W/Z$ is scale-invariant thanks to the dilaton factor $\chi$, and so is the mass of the SM fermions $f$: $g_Y  \bar f h f
 =(g_Y v/\sqrt{2}) (\chi \bar f f)$, all with the scale dimension 4.
 }
This yields the mass of the (pseudo-)dilaton as the Higgs $M_\phi^2=2\lambda v^2$, which is in accord  with the Partially Conserved Dilatation Current (PCDC):
\begin{equation}
M_\phi^2 F_\phi^2=-\langle0|\partial^\mu D_\mu|\phi\rangle F_\phi=-d_{\theta} \langle \theta_\mu^\mu\rangle =2\lambda v^4\langle \chi^2\rangle=2\lambda v^4\,,
\label{PCDC}
\end{equation}
with $F_\phi=v$, where $D_\mu$ is the dilatation current: $\langle 0|D_\mu(x) |\phi\rangle=-i q_\mu F_\phi e^{-i q x}$.  
 
Hence the SM Higgs
as it stands is a (pseudo) dilaton, with the {\it mass arising from the dimension 2 operator}
 in the potential, 
 \begin{equation}
 M_\phi^2=2\lambda v^2 
 \rightarrow 0
 \quad \left(\lambda\rightarrow 0\,, \,\,
 v=\sqrt{\frac{-\mu^2_0}{\lambda}} = 
 {\rm fixed}\,\ne 0\right)
 \label{conformallimit}
  \end{equation}
  (``conformal limit''\cite{Fukano:2015zua}).\footnote{This limit should be distinguished from the popular limit $\mu^2_0\rightarrow 0 $ with $\lambda=$fixed $\ne 0$, where the Coleman-Weinberg potential as the explicit scale symmetry breaking is generated by the trace anomaly  (dimension 4 operator) due to the quantum loop.
}
In fact the Higgs mass 125 GeV would imply $\lambda=M_\phi^2/(2 v^2) \simeq 1/8 \ll 1$.  It should be noted that {\it $\lambda\ll 1$ (with $v =$ fixed $\ne 0$) can be realized even when the underlying theory is strong coupling}, particularly when the {\it scale symmetry is operative},
 as we discuss below.

On the other hand, if we take the limit $\lambda \rightarrow \infty$, then the SM Higgs Lagrangian goes over to the  usual nonlinear sigma model {\it without scale symmetry}:  
\begin{equation}
{\cal L}_{{\rm NL}\sigma}=\frac{F_\pi^2}{4} {\rm tr} \left(\partial^\mu U\partial_\mu U^\dagger\right)\,,
\label{NLS}
\end{equation}
where the potential is decoupled and $\chi(x)\equiv 1$ is frozen, so that the scale symmetry breaking is transferred from the potential  to the kinetic term, 
which is no longer transform as the dimension 4 operator.  This is known to be a good effective theory (chiral perturbation theory) of the ordinary QCD which in fact lacks the scale symmetry at all,
perfectly consistent  with
the nonlinear sigma model, Eq.(\ref{NLS}). However, it cannot be true for  
the walking technicolor which does have the scale symmetry, and the effective theory must respect the symmetry of the underlying theory, in a form of the scale-invariant nonlinear sigma model Eq.(\ref{SNLSM}) in the conformal limit $\lambda\rightarrow 0$.   

Once rewritten in the form of Eq.(\ref{SNLSM}),  it is easy to see \cite{Fukano:2015zua}
that  the {\it SM Higgs Lagrangian is gauge equivalent to the ``scale-invariant HLS model'' (s-HLS)}\cite{Kurachi:2014qma}, a scale-invariant version of the HLS model \cite{Bando:1984ej,Bando:1987br,Harada:2003jx} \footnote{
The s-HLS model was also discussed in a different context, ordinary QCD in medium.\cite{Lee:2015qsa}
}
, which {\it contains massive spin-1 states}, spontaneously broken HLS gauge bosons, as  possible yet other composite states  in some
underlying theory hidden behind the 
SM Higgs:
\begin{eqnarray}
{\cal L}_{\rm s-HLS} &=& {\cal L}_{\rm Higgs}+ {\cal L}_{\rm Kinetic} \left(V_\mu\right)\,,\nonumber \\
{\cal L}_{\rm Higgs}&=& \chi^2 \cdot \left(\frac{1}{2} \left(\partial_\mu \phi\right)^2 + {\cal L}_A+ a {\cal L}_V\right) -V(\phi) \,,
\label{SHLS}
\end{eqnarray} 
with $a$ being an arbitrary parameter, and  ${\cal L}_{\rm Kinetic} \left(V_\mu\right)$ is the kinetic term of the HLS gauge boson $V_\mu$
which is obviously scale-invariant. The mass term  of the HLS gauge bosons is given by $\chi^2 \cdot a {\cal L}_V= \chi^2\cdot a (F_\pi^2/4) {\rm tr} (g_{_{HLS}} V_\mu +\cdots)^2$, which yields the mass   $M_V^2=a g_{_{HLS}}^2 F_\pi^2$, ($F_\pi=v)$, where $g_{_{HLS}}$ is the gauge coupling of the HLS.  For the low energy $p^2<M_V^2$ where the kinetic term can be ignored, the HLS gauge boson $V_\mu$ becomes just an auxiliary field to be solved away to yield ${\cal L}_V=0$, and 
we are left with $ \chi^2 {\cal L}_A= \chi^2 \cdot (F_\pi^2/4) {\rm tr} (\partial_\mu U \partial^\mu U^\dagger)$. 
Hence ${\cal L}_{\rm s-HLS}$ is reduced back to the original SM Higgs Lagrangian ${\cal L}_{\rm Higgs}$ in nonlinear realization, Eq.(\ref{SNLSM}). Note that {\it the HLS gauge boson acquires the
scale-invariant mass thanks to the dilaton factor $\chi^2$}, the nonlinear realization of the scale symmetry, in sharp contrast to the {\it Higgs (dilaton)  which acquires mass only from the explicit breaking of the scale symmetry}. 

This form of the Lagrangian Eq.(\ref{SHLS}) is the same as that of the effective theory of the walking technicolor, except for the shape of the scale-violating 
potential $V(\phi)$  which has a scale dimension 4 (trace anomaly) in the case of the walking technicolor instead of  2 of the SM Higgs case (Lagrangian mass term). We shall come back to this later.

\section{Composite Higgs from the  NJL Model}

Let us now elaborate the composite Higgs model based on the strong coupling theory $G>G_{\rm cr}\ne 0$ pioneered by Nambu.
In the Nambu-Jona-Lasinio (NJL)  model \cite{Nambu:1961tp} for the $N_C-$component 2-flavored fermion $\psi$ the Lagrangian takes the form:
\begin{eqnarray}
{\cal L}_{\rm NJL} &=& \bar \psi i\gamma^\mu\partial_\mu \psi + \frac{G}{2} \left[ (\bar \psi \psi)^2 +(\bar \psi i \gamma_5 \tau^a \psi)^2\right]\nonumber\\
&=& \bar \psi \left(i\gamma^\mu\partial_\mu  +\hat \sigma +i\gamma_5 \tau^a \hat \pi_a \right)\psi -\frac{1}{2G} \left({\hat \sigma}^2 +{\hat \pi_a}^2\right)\,,
\end{eqnarray}
where equation of motion of the auxiliary fields  $\hat \sigma \sim G \bar \psi \psi$ and ${\hat \pi}^a \sim G \bar \psi i\gamma_5 \tau^a\psi$ are plugged back in the Lagrangian to get the original Lagrangian.
In the large $N_C$ limit ($N_C \rightarrow \infty$ with $N_C G\ne 0$ fixed), after rescaling the induced kinetic term to the canonical one, $Z_\phi^{1/2}  \hat \sigma \rightarrow \hat \sigma$,
 the quantum theory for $\hat \sigma$ and $\hat \pi$ sector  yields 
precisely the same form as the SM Higgs Eq.(\ref{Lag:sigma}), 
with
\begin{eqnarray}
\mu_0^2&=& \left(\frac{1}{G} - \frac{1}{G_{\rm cr}}\right) Z_\phi^{-1} =-2m_F^2=- v^2 Z_\phi^{-1} =- \lambda v^2 < 0 \quad (G>G_{\rm cr}=\frac{4\pi^2}{N_C \Lambda^2} )
\nonumber\\
\lambda&=& Z_\phi Z_\phi^{-2} = Z_\phi^{-1} = \left[\frac{N_C}{8\pi^2} \ln \frac{\Lambda^2}{m_F^2}\right]^{-1} \sim \left[\frac{N_C}{8\pi^2} \ln \frac{\Lambda^2}{v^2}\right]^{-1} \,,
\label{tachyon}
\end{eqnarray}
where the gap equation has been used:
\begin{equation}
\frac{1}{G} - \frac{1}{G_{\rm cr}}= -\frac{N_C}{4\pi^2} m_F^2 \ln \frac{\Lambda^2}{m_F^2}=- 2 m_F^2 Z_\phi =-F_\pi^2=-v^2\,.
\label{gap}
\end{equation}

Eq.(\ref{tachyon}) shows that the {\it tachyon with $\mu^2<0$ is in fact generated} by the dynamical effects  for the {\it strong coupling}  $G>G_{\rm cr}\ne 0$,
 corresponding to the generation of mass $m_F\ne0$ in the gap equation.
Or, we can explicitly see it by computing the $\bar \psi \psi$ bound state  using the  $m_F=0$  solution  (wrong solution) of the gap equation at $G>G_{\rm cr}$. The correct spectrum $M_{\pi}^2=0, M_{\sigma}^2 =2 \lambda v^2 =-2\mu_0^2=4m_F^2$ can be 
obtained when we use the correct solution $m_F\ne 0$ in the gap equation. 
The last equality $M_\sigma^2=4 m_F^2$ is specific to the $N_C \rightarrow \infty$ (with $N_C G\ne 0$ fixed) limit of the NJL model ({\it ``weak coupling'' limit}  $G >G_{\rm cr} \sim 1/N_C\rightarrow 0$ in the {\it strong coupling phase}),  but not the general outcome of the NJL model nor the generic  linear sigma model.

There are two extreme limits for $\lambda$ in Eq.(\ref{tachyon}) : 
$\lambda \rightarrow0$ ($N_C\gg 1$ and/or $\Lambda/v^2 \gg 1$) reproduces precisely the conformal limit, or scale-invariant nonlinear sigma model limit, Eq.(\ref{SNLSM}), of the SM Higgs Lagrangian, 
while $\lambda\rightarrow \infty$ ($N_C, \Lambda^2/v^2 ={\cal O}(1)$) does  the nonlinear sigma model limit without scale symmetry, Eq.(\ref{NLS}).

We are interested in the limit 
$\lambda =[N_C \ln (\Lambda^2/v^2)/(8\pi^2)]^{-1} \rightarrow 0$ (conformal limit in Eq.(\ref{conformallimit}))
realized for $\Lambda/v\rightarrow \infty$ and/or $N_C\rightarrow \infty$, with $v=F_\pi=F_\phi \ne 0$ fixed.\footnote{
If $\Lambda$ is regarded as a physical cutoff in contrast to the nonperturbative renormalization arguments below, this argument would not be realistic for the 125 GeV Higgs with $\lambda\simeq 1/8$, corresponding to $\Lambda\simeq v\cdot e^{32\pi^2/N_C} \gg 10^{19}$ GeV. For the NJL model with $N_D$ doublets, however, we would have $\Lambda\simeq v \cdot e^{32\pi^2/(N_D N_C)} \sim 10^{11}$ GeV for $N_D=N_C=4$.
}
 Then the effective Lagrangian in the large $N_C$ limit takes precisely the same as the SM Higgs Lagrangian, 
which is further equivalent to the scale-invariant nonlinear sigma model, Eq.(\ref{SNLSM}), as mentioned before. Now the SM
Higgs is identified with the composite (pseudo-)dilaton  
with mass vanishing  $M_\phi^2 =2 \lambda v^2 \rightarrow 0$.

The limit theory gives an {\it interacting low energy effective theory even in the $\Lambda/v \rightarrow \infty$ limit}: a scale-invariant nonlinear sigma model ~\cite{Matsuzaki:2012mk,Matsuzaki:2015sya} where massless $\pi$ and $\phi$ are {\it interacting each other} with 
the (derivative) couplings $\sim (1/F_\pi, 1/F_\phi) \ne 0$.  It is actually the basis for the {\it scale-invariant chiral perturbation theory} (sChPT) with the derivative expansion as a loop expansion \cite{Matsuzaki:2013eva}, although the Yukawa couplings of
$\pi,\phi$ to the fermions are vanishing $g_Y\sim m_F/F_\pi, m_F/F_\phi \rightarrow 0$ (The composite particles are still interacting due to the loop divergence compensation of the vanishing Yukawa coupling). 
This limit should be sharply distinguished from a similar limit $\Lambda/m_F \rightarrow \infty$, $m_F=$fixed (not $\Lambda/v \rightarrow \infty$, $v=$fixed), which is  the famous  triviality limit (Gaussian fixed point) where the theory becomes a free theory: free massive scalar for $G<G_{\rm cr}$ and free tachyon for $G>G_{\rm cr}$, with not just the Yukawa couplings but all the couplings vanishing. 

One might wonder why dilaton in NJL model? Obviously the NJL model has  the explicit scale-breaking coupling $G$ having dimension $[M]^{-2}$. But this scale is an ultraviolet scale to which the low energy effective theory
is insensitive. This is in exactly the same sense as in the walking technicolor where the intrinsic scale $\Lambda_{\rm TC}$ generated by the trace anomaly can be far bigger than the infrared scale of spontaneous
symmetry breaking $F_\pi, F_\phi ={\cal O}  (v) \ll \Lambda_{\rm TC}$ thanks to the approximate scale symmetry due to the almost nonnruning coupling.  

Actually we can formulate the nonperturbative running of the (dimensionless) four-fermion coupling $g=\frac{N_C\Lambda^2}{4\pi^2} G$ in the same way as the Miransky nonperturbative renormalization:\footnote{
The argument here is somewhat similar to the renormalizability arguments of the $D$-dimensional NJL model ($2<D<4$) \cite{Kikukawa:1989fw}
  and the gauged NJL model \cite{Kondo:1991yk}, although the explicit scale-breaking from the Lagrangian parameters, i.e., the  four-fermion interaction and fermion mass term (if present), depend on the renormalization point (vanish at the UV limit).
}
The gap equation Eq. (\ref{gap}) reads
\begin{equation}
\left(\frac{1}{g_{\rm cr}}  - \frac{1}{g}\right) \Lambda^2 = m_F^2 \ln \frac{\Lambda^2}{m_F^2} \simeq \frac{4\pi^2}{N_C}v^2\,,\quad g_{\rm cr}=1\,,
\label{gap2}
\end{equation}
which leads to  a nonperturbative beta function 
\begin{equation}
\beta (g) = \Lambda \frac{\partial g(\Lambda)}{\partial \Lambda}\Bigg|_{v=\rm fixed}=-\frac{2}{g_{\rm cr}} g\cdot  (g-g_{\rm cr})\,,\quad g(\mu) = g_{\rm cr}\frac{1}{
1-\frac{4\pi^2} {N_C g_{\rm cr}}\frac{v^2}{\mu^2} 
}
\end{equation}
by {\it fixing $v=$ constant} and taking $\Lambda \rightarrow \infty$. Thus $g=g_{\rm cr}=1$ is the ultraviolet fixed point 
that the running coupling $g(\mu)$ 
reaches even much faster than the walking coupling in Eq.(\ref{NPrun}). 
Then  the scale symmetry is operative 
$g(\mu) \approx g_{\rm cr}$ for 
the wide region $\frac{4\pi^2}{N_C g_{\rm cr}} v^2< \mu^2< \Lambda^2$, with the {\it explicit scale symmetry breaking in the dimension 2 operator}: $\theta_\mu^\mu= \frac{\beta(g)}{g} \frac{G}{2} \left[
\left(\bar \psi \psi\right)^2 +\left(\bar \psi i\gamma_5 \tau^a \psi \right)^2
\right]= -\lambda v^4 \chi^2$, \footnote{
 Note that $- \langle \bar \psi_i \psi_j \rangle = \delta_{i,j} \Lambda^2 m_F N_C/(4\pi^2)= Z_m^{-1}\,\delta_{i,j} v^3 N_C /(4\pi^2) $, where $Z_m^{-1}=Z_m^{-1}(\Lambda/v)= (\Lambda/v)^2 [N_C \ln (\Lambda^2/v^2)/(4\pi^2)]^{-1/2} $ is the mass renormalization constant. 
This implies  $\gamma_m = Z_m \Lambda \frac{\partial Z_m^{-1}}{\partial \Lambda} =2 - 1/\ln(\Lambda^2/v^2)
\rightarrow 2$, and hence the operators have  scale dimension $d_{\bar \psi \psi}= 1$ and $d_{(\bar \psi \psi)^2}= 2$.
Thus we may write $\bar \psi_i \psi_j= -Z_m^{-1}\delta_{i,j} v^3 N_C/(4\pi^2) \cdot \chi $, or $(G/2)(\bar \psi \psi)^2=2 g \Lambda^2 v^2\cdot \chi^2/[\ln(\Lambda^2/v^2)]$. 
The gap equation implies $\beta(g)/g = -g (4\pi^2/N_C) (v^2/\Lambda^2)$.
Putting all together, we have $\beta(g)/g\cdot (G/2) \cdot
(\bar \psi \psi)^2|_{g\rightarrow g_{\rm cr}=1}= - \lambda v^4\chi^2$. 
}
where $\lambda = 8\pi^2/[N_C\ln(\Lambda^2/v^2)]\rightarrow 0$ as in Eq.(\ref{tachyon}).
The PCDC follows precisely the same way as in the SM Higgs as $M_\phi^2 F_\phi^2 =-d_{(\bar \psi \psi)^2} \langle \theta_\mu^\mu\rangle =2 \lambda v^4$ (See below Eq.(\ref{SNLSM})).
In any case the trace of energy-momentum tensor vanishes in the limit $\lambda \sim 1/[N_C\ln (\Lambda^2/v^2)] \rightarrow 0$, and  the dilaton mass should come from the trace anomaly in the $1/N_C$ sub-leading loop effects, or the chiral loops of the effective theory Eq.(\ref{SNLSM}).  

Again the spin 1 composites can also be introduced via HLS, precisely in the same way as Eq.(\ref{SHLS})  for the SM Higgs 
Lagrangian. This time it can be done more explicitly by introducing the vector/axialvector type four-fermion coupling which
in fact become the ``explicit'' composite HLS gauge bosons.(See section 5.3 of Ref.\cite{Bando:1987br}).

One of the concrete composite Higgs models as the 
straightforward
application of the NJL type theory is the top quark condensate model (Top-Mode Standard Model) \cite{Miransky:1988xi,Nambu:1989jt,Bardeen:1989ds}. The crucial ingredient of the model is again the {\it non-zero critical coupling}  {\it in sharp contrast to the weakly-coupled BCS theory which has $g_{\rm cr}=0$} as already mentioned.:
{\it only the top quark coupling is strong coupling larger than the critical coupling}
$G_t > G_{\rm cr}$ while others are less, $G_{b,c,s,d,u} <G_{\rm cr}$, so that only the top acquires the dynamical mass of order of weak scale  ${\cal O}(v)$ to produce
only three NG bosons to be absorbed into the $W/Z$ bosons \cite{Miransky:1988xi,Bardeen:1989ds}. The large $N_C$ limit relation $M_\phi=2 m_t$ is modified by the effects of the SM gauge interactions,
$M_\phi \simeq \sqrt{2} m_t$ \cite{Bardeen:1989ds, Shuto:1989te},  and further reduced 
by the non-leading order in $1/N_C$ expansion \cite{Bardeen:1989ds}. Different reductions have been considered,  top seesaw \cite{Dobrescu:1997nm}, and its NG boson Higgs version \cite{Fukano:2013aea}.  Updated detailed discussions are given in the talk by H. Fukano in this meeting.\cite{Fukano:2015ysa}

\section{Composite Higgs in Walking Technicolor: Technidilaton as the 125 GeV Higgs}

We have already seen that the SM Higgs is a dilaton in the conformal limit.  Here I come to the main subject,
the  technidilaton in the Walking Technicolor, the SCGT. Details are given in the talk by S. Matsuzaki in this meeting.\cite{Matsuzaki:2015pda}

Let us discuss a typical walking technicolor, the QCD-like vector-like $SU(N_C)$ gauge theory with $N_F$ massless technifermions,
particularly near the 
``anti-Veneziano limit'' (in distinction to the original Veneziano limit with $N_F/N_C\ll 1$):\cite{Matsuzaki:2015sya}
\begin{equation} 
N_C \rightarrow \infty \quad {\rm and} \quad \lambda\equiv N_C \cdot \alpha = {\rm fixed},
 \quad  {\rm with} \quad r\equiv N_F/N_C ={\rm fixed} \,\, \gg 1\,.
 \label{antiVeneziano}
\end{equation}

Such a limit is ideal  for the
ladder approximation with nonrunning coupling, since the coupling growth in the infrared region by the (anti-screening) gluon loops is cancelled by the opposite (screening) effects of the increasing fermion loops as $N_F$ increases, and eventually
levels off by balance at certain large number $N_F> N_F^* (\gg 2)$ ($N_F<11N_C/2$), as realized by  the 
 infrared (IR) fixed point $\alpha_*$ such that  $\beta^{({\rm perturbative)}}(\alpha=\alpha_*)=0$ as demonstrated in the two-loop beta function  (Caswell-Banks-Zaks (CBZ) IR fixed point) \cite{Caswell:1974gg}, with $N_F^*\simeq 8 (N_C=3)$.   Thus  
the input perturbative coupling in the SD equation becomes almost nonrunning,
$\alpha(\mu^2) \simeq \alpha_*$,
in the infrared region $\mu^2< \Lambda_{\rm TC}^2$  (infrared conformality), where $\Lambda_{\rm TC}$ is the intrinsic scale, analogous to the $\Lambda_{\rm QCD}$, generated by the perturbative trace anomaly 
 responsible for the perturbative (asymptotically-free) running of the coupling in the UV region $\mu^2>\Lambda_{\rm TC}^2$.  It plays the role of the cutoff $\Lambda$ in the ladder approximation.
 
Then the anti-Veneziano limit Eq.(\ref{antiVeneziano}) corresponds to the original walking technicolor \cite{Yamawaki:1985zg,Bando:1986bg}  based on the ladder SD equation in Landau gauge for the fermion propagator $ i S_F^{-1}(p)= \gamma^\mu p_\mu -\Sigma (-p^2)$, with the gauge coupling $\alpha(p^2)\equiv g(p^2)^2/(4\pi^2)\equiv \alpha=$constant for $|p^2|<\Lambda^2=\Lambda^2_{\rm TC}$, which has the broken solution, 
$\Sigma(p^2=-m_F^2) =m_F\ne 0$, in the strong coupling phase
$C_2 \alpha_*>C_2 \alpha_{\rm cr}=\pi/3\ne 0$, where $C_2=(N_C^2-1)/(2N_C)$ is the quadratic Casimir. All the ladder results are intact in the limit:  Eqs.(\ref{Miranskys}-\ref{gamma} ) and 
the technidilaton as a composite Higgs. While in the weak coupling phase (``conformal window'') $C_2 \alpha_*<C_2 \alpha_{\rm cr}$, there remains the unbroken approximate scale symmetry, and no bound states exist
(``unparticle''). 

 When $\alpha\simeq \alpha_*>\alpha_{\rm cr}$, such that $N_F >N_F^{\rm cr} (\simeq 4 N_C >N_F^*)$\cite{Appelquist:1996dq}, the SD equation has spontaneous breaking solution $m_F\ne 0$, arising from the technifermion condensate $\langle \bar F F\rangle\ne 0$, 
 which obviously {\it breaks both chiral symmetry and the scale symmetry spontaneously}.
  The scale symmetry is also {\it broken explicitly  by the same origin $m_F\ne 0$}:  
  the would-be IR fixed point actually is washed out, with the coupling starting running (walking)  as in Eq.(\ref{NPrun}),
  according to 
  the nonpertubative beta function Eq.(\ref{NPbeta}) responsible for a tiny  {\it nonperturbative trace anomaly of dimension 4 operator}: $\theta_\mu^\mu= \beta^{(\rm NP)}(\alpha)/(4\alpha) \cdot G_{\mu\nu}^2 =  {\cal O}(m_F^4) (\ll \Lambda_{\rm TC}^4)$, where  $G_{\mu\nu}$ is the
technigluon field strength with $G_{\mu\nu}^2$ also induced by the $m_F$. (Usual (perturbative) trace anomaly corresponding to the asymptotically free running of order ${\cal O} (\Lambda_{\rm TC}^4)$, has been subtracted out.)
  In accord with the smallness of the trace anomaly, the coupling is still almost nonrunning as in Eq.(\ref{NPrun}) for a wide
  infrared region $m_F<\mu<\Lambda_{\rm TC}\, (m_F\ll \Lambda_{\rm TC})$. 
  
\begin{figure}[h]
\begin{center}
\includegraphics[width=5.5cm]{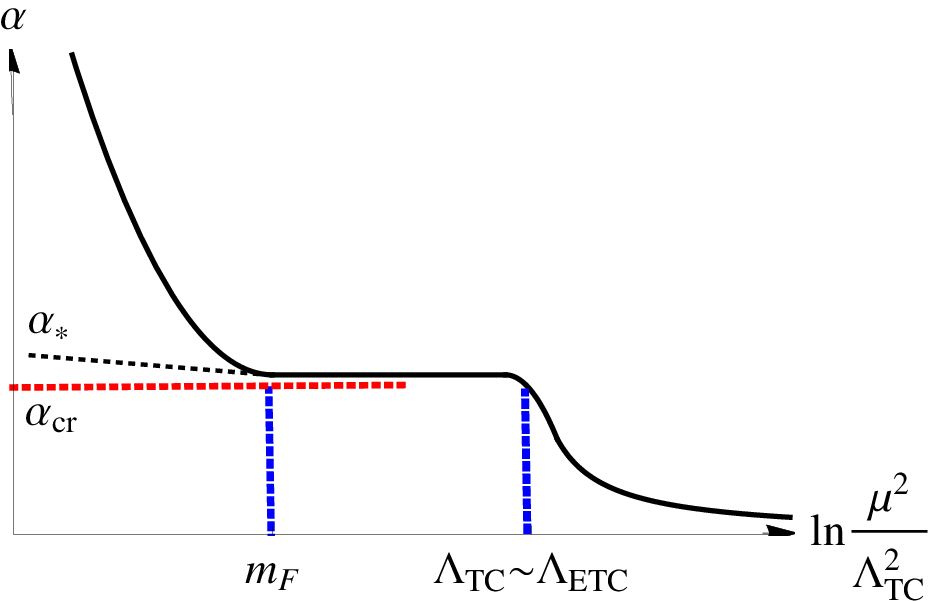} 
\hspace{15pt}
   \includegraphics[width=5.5cm]{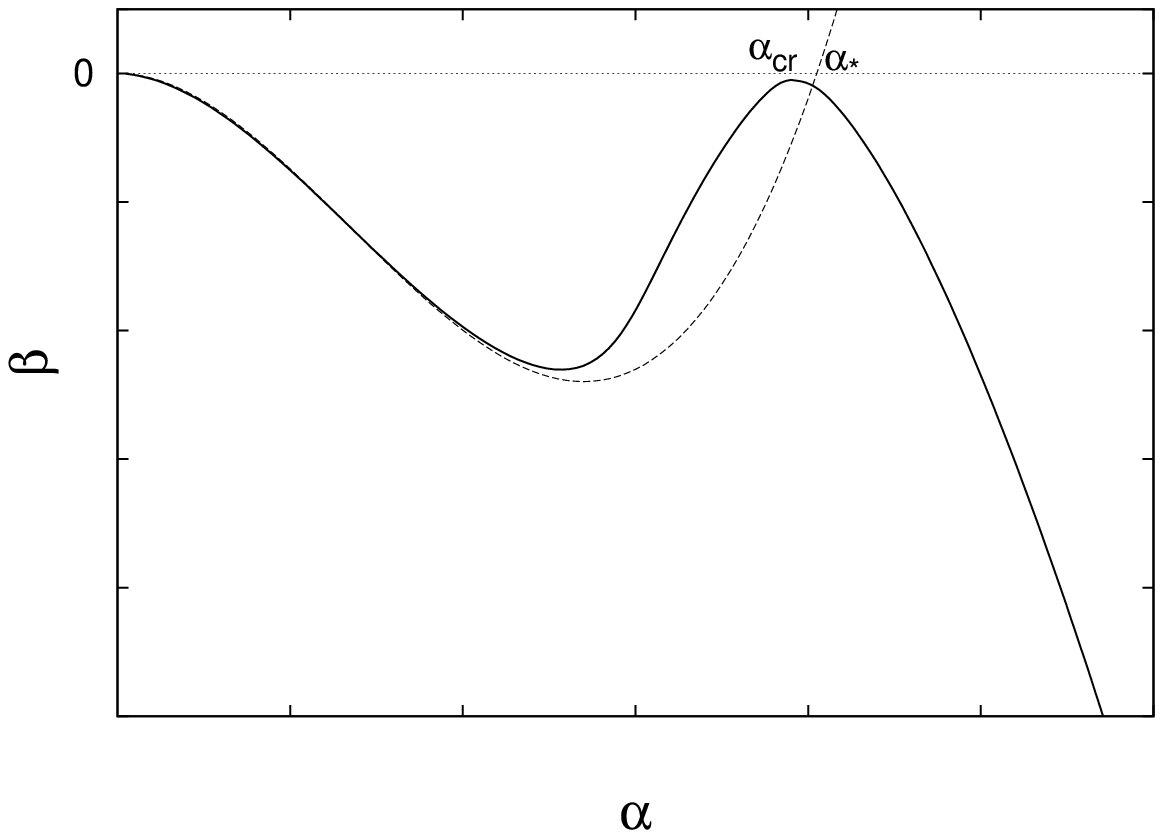} 
\vspace{15pt}
\caption{ Possible perturbative running coupling (left) and the beta function (right) in the region  $\alpha<\alpha_{\rm cr}$, in comparison with the nonperturbative region  $\alpha>\alpha_{\rm cr}$.
}
\label{beta:whole}
\end{center} 
 \end{figure}

  Note \cite{Matsuzaki:2015sya} that Eq.(\ref{NPbeta}) has a multiple zero at $\alpha=\alpha_{\rm cr}\simeq \alpha_*$ (not linear zero) and 
   is completely different from the two-loop beta function having the CBZ IR fixed point (having linear zero at $\alpha_*$) which  
 is no longer valid in the broken phase $\alpha>\alpha_*\simeq \alpha_{\rm cr}$, 
where $\alpha (\mu)\searrow \alpha_{\rm cr}$ ($\mu \nearrow$).
Then the would-be {\it IR fixed point $\alpha_* \simeq \alpha_{\rm cr}$ is also regarded as 
the UV fixed point} of the nonperturbative running (walking) coupling $\alpha(\mu) \approx \alpha_{\rm cr}$~\cite{Yamawaki:1985zg} 
in the wide IR region $m_F < \mu < \Lambda=\Lambda_{\rm TC}$ for the characteristic large hierarchy  $m_F \ll \Lambda_{\rm TC}$\cite{Yamawaki:2007zz,Hashimoto:2010nw}.  
(See also Ref.\cite{Kaplan:2009kr} for a similar observation.)  See Fig. \ref{beta:whole}.\footnote{
 Note also that the walking technicolor in the UV region $\mu^2>\Lambda_{\rm TC}^2$ must be changed into only a part of some larger picture, such as the Extended Technicolor (ETC)\cite{Dimopoulos:1979es} or models having both technifermions and SM fermions as composites on the equal footing \cite{Yamawaki:1982tg}, to provide the mass to the SM fermions through communicating technifermion condensate
to the SM fermions, so that discussing the walking technicolor in isolation  does not make sense for $\mu^2 >\Lambda_{\rm TC}^2$.  
}

Now we come to our core result,  ladder evaluation of  the nonperturbative trace anomaly in the anti-Veneziano limit, which then yields the mass $M_\phi$ and decay constant $F_\phi$ of the technidilaton $\phi$ 
through PCDC \cite{Bando:1986bg}  as in Eq.({\ref{PCDC}):\cite{Matsuzaki:2015sya,Hashimoto:2010nw}  
 \begin{eqnarray}
 M_\phi^2 F_\phi^2&=&  
-d_\theta \langle \theta_\mu^\mu \rangle 
=-  \frac{\beta^{(\rm NP)}(\alpha (\mu^2))}{\alpha(\mu^2)}
\, \langle G_{\mu \nu}^2(\mu^2)\rangle 
\simeq N_C N_F\frac{16 }{\pi^4} m_F^4\quad \left(d_\theta=4 \right)
\label{PCDC2}\\
&\simeq& 2.5\left[\frac{8}{N_F}\frac{4}{N_C}\right] v^4 \,. \quad \left(v= 246\,{\rm GeV}\right)
\label{PCDC3}
\end{eqnarray}

Firstly, the rightmost value in Eq.(\ref{PCDC2}) can be obtained by two different ladder calculations: 
one through direct evaluation of the vacuum energy by the effective potential at the stationary point (Solution of the SD equation, $\Sigma=\Sigma_{\rm sol}$)\cite{Gusynin:1987em}, $E=V_{\rm eff} (\Sigma =\Sigma_{\rm sol}) =\langle\theta^0_0\rangle
=(1/4)\langle\theta^\mu_\mu\rangle$,  the other  through the ladder evaluation  of the trace anomaly \cite{Hashimoto:2010nw,Matsuzaki:2015sya}, i.e., the  technigluon condensate $\langle G_{\mu\nu}^2\rangle$ times the nonperturbative beta function
Eq.(\ref{NPbeta}), both in precise agreement with each other. The agreement is  in highly nontrivial manner, being  {\it independent of the renormalization point $\mu$} as it should be:
 $\langle G_{\mu \nu}^2(\mu^2)\rangle \sim \ln^3 (\mu^2/m_F^2)$,  while $\beta^{({\rm NP})}  (\alpha (\mu^2)) /\alpha(\mu^2) \sim  1/\ln^3 (\mu^2/m_F^2)$, precisely cancelled by each other.\cite{Matsuzaki:2015sya}
 
Secondly, Eq.(\ref{PCDC3}) is obtained by use of the Pagels-Stokar formula:
\begin{equation}
 v^2=(246\,  {\rm GeV})^2 
= N_D F_\pi^2 \simeq 
N_F N_C\frac{1}{4\pi^2} \, m_F^2   
\simeq  m_F^2 \left[\frac{N_F}{8}\frac{N_C}{4}\right],
\end{equation}
and the result indicates {\it important $N_F,N_C-$ dependence of $M_\phi^2F_\phi^2$ in the anti-Veneziano limit when $v=$ fixed}\cite{Matsuzaki:2015sya}.
Since the technidilaton is a flavor-singlet bound state, its decay constant by definition scales like $F_\phi^2 \propto N_F N_C m_F^2 (\propto v^2)$ (Actually $F_\phi^2 \simeq N_F N_C m_F^2$). Then 
 $M_\phi^2/F_\phi^2, M_\phi^2/v^2 \sim 1/(N_F N_C) \rightarrow 0$ in the anti-Veneziano limit, where the technidilaton becomes NG boson although no exact
 massless limit exists: the situation is  in the same sense as the 
 $\eta^\prime$ meson in the original Veneziano limit $N_C \rightarrow \infty$ with $N_C \alpha=$fixed, and $N_F/N_C \ll 1$.\footnote{
 There exists no exact massless limit in the conformal phase transition at $\alpha=\alpha_{\rm cr}$, with $m_F=0$, where no massless spectrum exists (conformal phase),
 in sharp contrast to the Ginzburg-Landau phase transition where the spectrum continuously passes through the phase transition point with massless particles. \cite{Miransky:1996pd} 
 }
 
 Numerically, Eq.(\ref{PCDC3}) implies that
 \begin{equation}
 F_\phi\simeq 5 \,v \quad {\rm for} \,\, M_\phi \simeq \frac{v}{2} \simeq 125\,{\rm GeV}\quad \left(N_F=8, N_C=4\right)\,,
 \end{equation} 
in the one-family model, which is best fit to the current LHC data of the 125 GeV Higgs. \cite{Matsuzaki:2012mk,Matsuzaki:2015sya}
Similar results are also obtained in the holographic model for the walking technicolor.\cite{Matsuzaki:2012xx}

\section{Discovering Walking Technicolor at LHC} 

Now to the LHC phenomenology of the walking technicolor. Details are given in the talk by S. Matsuzaki in this meeting.\cite{Matsuzaki:2015pda}
The model
has in general a larger chiral symmetry $SU(N_F)_L\times SU(N_F)_R$ $(N_F>2)$ spontaneously broken  by the technifermion condensate $\langle \bar F F \rangle \ne 0$, or
the technifermion dynamical mass $m_F\ne 0$, down to the diagonal $SU(N_F)_V$. Also {\it spontaneously broken by the same 
technifermion condensate} is the {\it approximate scale symmetry} due to the almost nonruning (walking) coupling $\alpha(\mu^2) \approx \alpha_{\rm cr}$ in the wide infrared region $m_F^2 < \mu^2<\Lambda_{\rm TC}^2$ ($m_F\ll \Lambda_{\rm TC}^2$).
Moreover the scale symmetry is {\it simultaneously broken explicitly  by the same origin $m_F\ne 0$}, an emergent infrared mass scale, resulting in
the 
{\it nonperturbative trace anomaly of dimension 4 operator} as was given in Eq.(\ref{PCDC2}). 

The effective theory  should have the same symmetry structure as the underlying theory.
Namely, the {\it nonlinear realization of both the chiral symmetry and the scale symmetry}, which implies the {\it scale-invariant nonlinear sigma model}~\cite{Matsuzaki:2012mk,Matsuzaki:2015sya}. 
Then the effective theory of the walking technicolor with $N_F$ massless flavors takes {\it precisely the same scale-invariant form as the nonlinearly realized SM Higgs Lagrangian} in Eq.(\ref{SNLSM}), with $U=e^{i \pi^a\, T^a}$ being $N_F\times N_F$ unitary matrix (${\rm tr} \,T^a=0\,, {\rm tr} (T^aT^b)=\delta^{ab}/2,\,a=1,\cdots, N_F^2-1$), except that 
the explicit scale breaking comes from the different potential $V^{(4)}(\phi)$~\footnote{
This potential is indeed obtained by the explicit ladder computation of the effective potential  at the conformal phase transition point: $V^{(4)}(\phi)= -( 4N_F N_C m_F^4/\pi^4) \chi^4 (\ln \chi-1/4)$, in precise agreement with
Eq.(\ref{WTC}) through Eq.(\ref{PCDC2}). See Eq.(65) in Ref.~\cite{Miransky:1996pd}} 
 responsible for the {\it trace anomaly of dimension 4 operator} this time:~\cite{Matsuzaki:2012mk,Matsuzaki:2015sya}
 \begin{eqnarray}
  {\cal L}_{\rm WTC}
 &=&\chi^2 \cdot \left[ \frac{1}{2} \left(\partial_\mu \phi\right)^2  +\frac{F_\pi^2}{4}{\rm tr} \left({\cal D}_\mu U {\cal D}^\mu U^\dagger\right)\right] -V^{(4)}(\phi)-V^{(\rm SM)}(\phi) \nonumber \,,\\
 V^{(4)}(\phi)&=&-\ln \chi\cdot \frac{\beta^{(\rm NP)}(\alpha)}{4\alpha} G_{\mu\nu}^2= \frac{M_\phi^2 F_\phi^2}{4} \chi^4 \left(\ln\chi-\frac{1}{4}\right)\nonumber\\
V^{(\rm SM)}(\phi)&=& - \chi^{2-\gamma_m} \left(m_f \chi \bar f f\right)  -\ln \chi \left[\frac{\beta_F(\alpha_s)}{4\alpha_s} {G^{({\rm gluon})}_{\mu\nu}}^2 
+ \frac{\beta_F(\alpha_e)}{4\alpha_e} {F^{({\gamma})}_{\mu\nu}}^2\right]\,,\,\nonumber\\
 \chi &= &\exp \left(\frac{\phi}{F_\phi}\right)\,,
 \label{WTC}
 \end{eqnarray} 
where {\it $F_\phi\ne F_\pi=v/\sqrt{N_D}=v/\sqrt{N_F/2}$ in general} in contrast to the SM Higgs case $F_\phi=F_\pi=v$, the electroweak gauging was done as usual  $\partial_\mu U\Rightarrow {\cal D}_\mu U= \partial_\mu U -i g_2 W_\mu U +i g_Y U B_\mu$, and we have
added  $V^{(\rm SM)}(\phi)$, the scale symmetry breaking terms related  to the SM particles arising from the technifermion contributions: mass term of the SM fermion $f$,  (one loop)  technifermion  contributions to the 
trace anomaly for  the gluon and photon operators, with  $\beta_F(g_s) = \frac{g_s^3}{(4\pi)^2} \frac{4}{3} N_C$ and  $\beta_F(e) = \frac{e^3}{(4\pi)^2} \frac{16}{9} N_C$. 
It is obvious that ${\theta_\mu^\mu}^{({\rm TC})}=- \delta_D V^{(4)}(\phi) = \beta^{(\rm NP)}(\alpha)/(4\alpha)\cdot G_{\mu\nu}^2=- (M_\phi^2F_\phi^2/4) \chi^4$ up to total derivative, corresponding to
the PCDC with $d_\theta=4$ ($\langle \chi\rangle=1$), Eq.(\ref{PCDC2}). 

The technidilaton potential   $V^{(4)}(\phi)$ is expanded in $\phi/F_\phi$:
\begin{equation} 
V^{(4)}(\phi)
= - \frac{M_\phi^2 F_\phi^2}{16} +\frac{1}{2}M_\phi^2\,\phi^2 +\frac{4}{3} \frac{M_\phi^2}{F_\phi} \,\phi^3 
+ 2 \frac{M_\phi^2}{F_\phi^2}\, \phi^4 
+ \cdots 
\,,
\label{dilatonpotential}
\end{equation}
which shows a remarkable fact that in the anti-Veneziano limit 
the technidilaton  self couplings (trilinear and quartic couplings) are highly suppressed:
$
\lambda_{\phi^3}=4M_\phi^2/(3F_\phi) \sim 1/\sqrt{N_F N_C}$,
$\lambda_{\phi^4}=2 M_\phi^2/F_\phi^2 \sim 1/(N_F N_C)
$, 
by $M_\phi/F_\phi \sim 1/\sqrt{N_F N_C}$ and $M_\phi \sim N_F^0 N_C^0$. 
Numerically, we may compare the technidilaton self couplings
with those
of the SM Higgs with $m_h=M_\phi=125$ GeV for $v/F_\phi \simeq 1/5$ in the one-family model ($N_F=8, N_C=4)$: \cite{Matsuzaki:2015sya}
\begin{eqnarray} 
\frac{\lambda_{\phi^3}}{\lambda_{h_{\rm SM}^3}}\Bigg|_{M_\phi=m_h} 
&=& \frac{\frac{4 M_\phi^2}{3 F_\phi}}{\frac{m_h^2}{2 v}} \Bigg|_{M_\phi=m_h}
\simeq \frac{8}{3} \left( \frac{v}{F_\phi}\right) \simeq 0.5 \,, \nonumber \\ 
\frac{\lambda_{\phi^4}}{\lambda_{h_{\rm SM}^4}} \Bigg|_{M_\phi = m_h} 
&=& \frac{\frac{2 M_\phi^2}{F_\phi^2}}{\frac{m_h^2}{8 v^2}}\Bigg|_{M_\phi=m_h} 
= 16 \left( \frac{v}{F_\phi} \right)^2 \simeq 0.6 
\,. 
\label{selfcouplings:0}
\end{eqnarray}
This shows that the {\it technidilaton self couplings, although generated by the strongly coupled interactions, are even smaller than those of the 
SM Higgs}, a salient feature of the approximate scale symmetry in the ant-Veneziano limit !!

The coupling of the technidilaton ($M_\phi=125$ GeV) to the SM particles
can be seen by expanding $\chi= 1+\phi/F_\phi +(1/2!)(\phi/F_\phi)^2+\cdots$ in Eq.(\ref{WTC}): 
\begin{eqnarray} 
  \frac{g_{\phi WW/ZZ}}{g_{ h_{\rm SM} WW/ZZ }}   
 =
 \frac{g_{\phi ff}}{g_{h_{\rm SM} ff}}  \,
 = 
  \frac{v}{F_\phi} 
  \,.  
  \label{WWZZ}
\end{eqnarray} 
 \begin{eqnarray} 
\frac{g_{\phi gg}}{g_{h_{\rm SM} gg}} 
\simeq 
\frac{v}{F_\phi} 
\cdot 
\left( 1 + 2 N_C \right),\,
\frac{g_{\phi \gamma\gamma}}{g_{h_{\rm SM} \gamma\gamma}} 
\simeq \frac{v}{F_\phi} 
\cdot 
 \left( \frac{63 -  16}{47} - \frac{32}{47} N_C \right)  
\,,  \label{g-dip-dig}
\end{eqnarray} 
where besides the technifermion, only the top and W of the SM contributions  were included at one-loop.
Note the couplings in Eq.(\ref{WWZZ}) with $v/F_\phi \sim1/5$ are even weaker than the SM Higgs, which are however compensated by
those in Eq.(\ref{g-dip-dig}) for  $gg$ and $\gamma\gamma$ rather enhanced
by the  extra  loop contributions of the technifermions other than the SM particles, particularly for large $N_C$, resulting in signal strength similar to the SM Higgs within the current experimental accuracy.  

In fact the current LHC data for 125 GeV Higgs are fit by the technidilaton as good as by the SM Higgs,
particularly for $N_F=8, N_C=4$, i.e.,  near the anti-Veneziano limit.~\cite{Matsuzaki:2012mk}  
Most recent detailed analyses are given in Ref. \cite{Matsuzaki:2015sya}.
It should be mentioned here that the one-family model will be most naturally imbedded into the ETC in the case for $N_C=4$ \cite{Kurachi:2015bva}.
More precise data at LHC Run II will discriminate among them, SM Higgs or technidilaton. We will see.

Next to the technipions: in the walking technicolor with $N_D =N_F/2>1$, the spontaneous breaking of the chiral symmetry larger than $SU(2)_L\times SU(2)_R$ produces NG bosons (technipions) more than 3 to
be absorbed into W/Z. Let us take the one-family model with $N_F=8$, which has colored techniquarks (3 weak doublets) $Q^a_i$ and non-colored technileptons (one weak-doublet) $L_i$
($a=1,2,3;i=1,2$), the resultant chiral symmetry being $SU(8)_L \times SU(8)_R$ \cite{Dimopoulos:1979sp}. There are 63 technipions, 60 of which are  unabsorbed  technipions.   All of them acquire the mass from the explicit chiral symmetry breaking due to the SM gauge and ETC gauge interactions. Due to the large anomalous dimension $\gamma_m\simeq 1$, the mass of them are all
enhanced to TeV region \cite{Kurachi:2014xla}, which will be discovered at LHC Run II.

Another signatures of the walking technicolor are higher resonances such as the spin 1 boson, the walking techni-$\rho$, 
walking techni-$a_1$, etc..
The straightforward $N_F$ extension of Eq.(\ref{SHLS})  is also obvious:  Eq.(\ref{WTC}) is gauge equivalent to the {\it scale-invariant HLS Lagrangian} explicitly constructed for one-family walking technicolor with $N_F=8$ \cite{Kurachi:2014qma}: 
\begin{eqnarray}
{\cal L}_{\rm s-HLS} &=& {\cal L}_{\rm WTC}+ {\cal L}_{\rm Kinetic} \left(V_\mu\right)\,,\nonumber \\
{\cal L}_{\rm WTC}&=& \chi^2 \cdot \left(\frac{1}{2} \left(\partial_\mu \phi\right)^2 + {\cal L}_A+ a {\cal L}_V\right) -V^{(4)}(\phi) \,,
\label{SHLS2}
\end{eqnarray} 
where the HLS gauge bosons $V_\mu$ in the mass term $ a \chi^2 {\cal L}_V=a e^{2\phi/F_\phi} {\rm tr} (g_{_{\rm HLS}} V_\mu +\cdots)^2$ are the bound states of the walking technicolor, the walking techni-$\rho$,
with {\it mass $M_V^2=a g_{_{\rm HLS}}^2 F_\pi^2$ being scale-invariant} thanks to the overall technidilaton factor $\chi^2$, as mentioned before. 
The loop expansion is formulated as the scale-invariant HLS perturbation theory\cite{Fukano:2015zua} in the same way as the scale-invariant chiral perturbation theory\cite{Matsuzaki:2013eva}, a straightforward extension of the (non-scale-invariant) HLS perturbation theory \cite{Harada:2003jx}.
The HLS is readily extendable to include techni-$a_1$, etc.\cite{Bando:1984ej,Bando:1987br,Harada:2003jx}, with a infinite set of the HLS tower being equivalent to the deconstructed extra dimension\cite{ArkaniHamed:2001ca} and/or the holographic models\cite{Son:2003et}, and the scale-invariant version of them are also straightforward and  {\it mass of all the higher HLS vector bosons
are scale-invariant}, an outstanding characterization in sharp contrast  to other formulations for the spin 1 bosons.  We will see at LHC Run II.

\section{Walking on the Lattice}
Finally, I briefly mention the walking technicolor on the lattice focussing on our own results by LatKMI.
Updated details are given in the talks by Y. Aoki, K-i. Nagai and H. Ohki in this meeting\cite{Aoki:2015aqa}.
 The LatKMI Collaboration started in 2010 for the lattice simulations on the possible candidate for the walking technicolor by 
systematic studies of the $N_F=16,12,8, 4$ degenerate fermions in $SU(3)$ gauge theories (dubbed ``Large $N_F$ QCD''), using the HISQ (Highly Improved Staggered Quarks) 
action with tree-level Symanzik gauge action. We have mainly focused on the low-lying fermionic bound states (plus some gluonic ones), i,e., pseudoscalar (denoted as $\pi$), scalar ($\sigma,a_0$), vector ($\rho$), axialvector mesons ($a_1$), and nucleon-like states ($N, N^*$), particularly the flavor-singlet scalar $\sigma$ as a candidate for the technidilaton.

We found \cite{Aoki:2012eq} that $N_F=12$ is consistent with the conformal window indicating no spontaneous chiral symmetry breaking, which is in agreement with the results of many other groups.  We further found \cite{Aoki:2013xza} that $N_F=8$ is consistent with the spontaneously broken
phase with remnants of the conformality with large anomalous dimension $\gamma_m \simeq 1$, namely the walking theory, in accord with other groups \cite{Appelquist:2014zsa}.

The most remarkable result of the LatKMI Collaboration is the {\it discovery of a light flavor-singlet scalar on the lattice} 
in both $N_F=12$ \cite{Aoki:2013zsa}  and $N_F=8$ \cite{Aoki:2014oha}. The $N_F=12$ results are also consistent with
other groups \cite{Fodor:2014pqa}.  Since $N_F=8$ seems to be a walking theory in the broken phase,  the light flavor-singlet scalar is particularly attractive as a candidate for the technidilaton.  Also $N_F=8$ is  of phenomenological 
relevance to the LHC data as well as direct relevance to the one-family model as the  most natural model building.
Future confirmation of our results is highly desired. Also $N_C=4$ simulations should be studied for various reasons as mentioned before.

I am really proud of the LatKMI Collaboration, although it will loose funding soon.

\section {Summary}
I have argued that the 125 GeV Higgs is a (pseudo-)dilaton even if it is described by  the Standard Model Higgs Lagrangian (!!)
which  is actually shown to be equivalent to the scale-invariant nonlinear sigma model with both chiral and scale symmetries 
are nonlinearly realized. The SM Higgs Lagrangian is further gauge equivalent to the scale-invariant Hidden Local Symmetry (HLS)
Lagrangian which include new massive vector bosons as the gauge bosons of the (spontaneously broken) HLS, 
with the mass being scale-invariant.  

All these features of the SM Higgs Lagrangian are reminiscence of the conformal UV completion behind the Higgs, the
existence of the underlying theory with (approximate) scale symmetry, so strong coupling as to produce composite states
such as the Higgs (dilaton), new vector bosons (HLS gauge bosons), etc.. We have seen that even the NJL model, though not gauge theory,
 can be regarded as such a
conformal UV completion. The walking technicolor, conformal SCGT,  is such a typical  
underlying theory, where the 125 GeV Higgs is a composite dilaton, the technidilaton.

The walking technicolor in the
anti-Veneziano limit  $N_C \rightarrow \infty$ with $N_C \alpha=$fixed $={\cal O}(1)$ and $N_F/N_C=$ fixed ($\gg 1)$ makes
the ladder approximation reasonable, which yields a naturally light and weakly coupled technidilaton through the PCDC:
 \begin{equation}
 M_\phi^2 F_\phi^2=  
-4 \langle \theta_\mu^\mu \rangle 
=-  \frac{\beta(\alpha (\mu^2))}{\alpha(\mu^2)}
\, \langle G_{\mu \nu}^2(\mu^2)\rangle 
\simeq N_C N_F\frac{16 }{\pi^4} m_F^4\,,
\end{equation}
independently of the renormalization point $\mu$,
where the scale symmetry is explicitly broken by the trace anomaly of the dimension 4 operator $G_{\mu\nu}^2$,
which is induced by $m_F$ the 
dynamical mass of the technifermion arising from the simultaneous spontaneous breaking of the scale symmetry and the chiral symmetry.

I have defined ``strong coupling theories'' as ``those having non-zero critical coupling'' $N_C g_{\rm cr}={\cal O}(1)$, even though its
value could be small $g \sim 1/N_C \ll 1$ in the typical large $N_C$ limit. The NJL model pioneered by Professor Nambu is 
the first and a typical example of such, to be distinguished from its preceding, the BCS theory, which has a zero critical coupling
$g_{\rm cr}=0$.  Existence of such a non-zero critical coupling in gauge theory 
was discovered by Maskawa and Nakajima in the
scale-invariant dynamics, ladder approximation, and became crucial for the walking technicolor with the coupling
$N_C \alpha> N_C \alpha_{\rm cr} ={\cal O}(1)$ in the spontaneous broken phase of the scale symmetry as well as the chiral symmetry. 

The existence of the non-zero critical coupling  is actually ``hidden'' even in the QCD which is regarded to have only one phase in the ordinary
situation without signal of the no-zero critical coupling: it manifests itself in  the extreme condition, such as the 
large number of fermions $N_F\gg N_C$ (so as to keep the asymptotic freedom), high temperature, high density, etc.. 

Indeed, it is the large $N_F$ QCD that models the walking technicolor where the large number of fermions give the screening effects and level off of
 the infrared coupling which otherwise brows up due to the gluon anti-screening effects (Caswell-Banks-Zaks infrared fixed point). 
 For large $N_F$ with the fixed point value smaller than the critical coupling,  the spontaneously broken  phase is gone (what we called conformal  phase transition). 
 Then the infrared scale invariance becomes manifest. 
 
 I would infer existence of the similar ``hidden'' nonzero critical coupling
  even for ordinary QCD at high temperature and/or high density, where the infrared coupling cannot blow up in the
 region below the temperature/density scale, effectively frozen in the infrared region (effective scale symmetry similar to the walking technicolor). 
 If the frozen coupling is smaller than the  hidden ``critical coupling'',
 then the spontaneously broken phase would be gone, resulting in  the quark-gluon plasma and/or the alternative 
 color superconductor (genuine BCS weak coupling for
 the Fermi surface (only for quark, not anti-quark)).  Then the effective scale invariance would  manifest itself.  
 
I am retiring from KMI, Nagoya University as of March 2015.
This meeting will probably be the last one of the series of the Nagoya SCGT workshops. I hope that  future of the Strong Coupling 
Gauge Theories 
will be fruitful ever, no matter how the SCGT workshop might be over. LHC and lattice activity will tell us something.
Thank you everybody, with my never-ending dream to toast to the old wine in a new bottle.
  
\section{Acknowledgements}
I would like to express my hearty thanks to all the participants in SCGT15 as well as those in the previous SCGT workshops.
This work was supported by the JSPS Grant-in-Aid for Scientific Research (S) \# 22224003 and (C) \#23540300.

\vspace{0.5cm}
[Note added]
After the SCGT15 workshop, there appeared an interesting report on the 2 TeV excess in the diboson channels \cite{Aad:2015owa}. 
We have argued \cite{Fukano:2015hga}
 that this would be 
a most natural candidate for the walking techni-$\rho$ as a gauge boson of the Hidden Local Symmetry described 
by the scale-invariant HLS model in Eq.(\ref{SHLS2}) \cite{Kurachi:2014qma}.  We further found \cite{Fukano:2015uga,Fukano:2015zua} 
that a salient feature of this possibility is
the scale symmetry which forbids the decay of the walking techni-$\rho$ to the 125 GeV Higgs (technidilaton) plus $W/Z$ (what we called
``conformal barrier''), in sharp contrast to the popular ``equivalence theorem''. This applies not only to the techni-$\rho$ but
also to  all the higher vector/axialvector resonances as the HLS gauge bosons, having scale-invariant mass. The LHC Run II will tell us whether or not it is the case.


\begin{thebibliography}{100}


\bibitem{Yamawaki:1985zg}
  K.~Yamawaki, M.~Bando and K.~Matumoto,
  Phys.\ Rev.\ Lett.\  {\bf 56}, 1335 (1986);
  M.~Bando, T.~Morozumi, H.~So and K.~Yamawaki,
  Phys.\ Rev.\ Lett.\ 
  {\bf 59}, 389 (1987).

\bibitem{Bando:1986bg}
  M.~Bando, K.~Matumoto and K.~Yamawaki,
  Phys.\ Lett.\  B {\bf 178}, 308 (1986). 
 

\bibitem{Miransky:1984ef}
  V.~A.~Miransky,
  Nuovo Cim.\  A {\bf 90} (1985), 149.

  \bibitem{Miransky:1996pd}
  V.~A.~Miransky and K.~Yamawaki,
  Phys. Rev. D {\bf 55}, 5051 (1997);
  Errata, {\bf 56}, 3768 (1997).

\bibitem{Maskawa:1974vs}
  T.~Maskawa and H.~Nakajima,
  Prog.\ Theor.\ Phys.\  {\bf 52} (1974), 1326;
 {\bf 54} (1975), 860.
  

\bibitem{Nambu:1961tp}
Y. Nambu, Superconductor Model of Elementary Particles and its Consequencies - Talk given at a conference at Purdue (1960);
  Y.~Nambu and G.~Jona-Lasinio,
  Phys.\ Rev.\  {\bf 122}, 345 (1961).


\bibitem{TC}
S.~Weinberg,
Phys.\ Rev.\ D {\bf 13} (1976), 974; 
D {\bf 19} (1979), 1277;
L.~Susskind,
Phys.\ Rev.\ D {\bf 20} (1979), 2619:
See for a review of earlier literature,
E.~Farhi and L.~Susskind,
Phys.\ Rep.\  {\bf 74} (1981), 277.

  
\bibitem{Leung:1985sn} 
  C.~N.~Leung, S.~T.~Love and W.~A.~Bardeen,
  Nucl.\ Phys.\ B {\bf 273}, 649 (1986).
  
\bibitem{Holdom:1981rm}
  B.~Holdom,
  Phys.\ Rev.\  D {\bf 24}, 1441 (1981).

 \bibitem{Akiba:1985rr}
T. Akiba and T. Yanagida,
Phys.\ Lett.\ B {\bf 169} (1986), 432;
T.~W. Appelquist, D. Karabali and L.~C.~R. Wijewardhana,
Phys.\ Rev.\ Lett.\  {\bf 57} (1986), 957; 
For an earlier work on this line see B. Holdom,
Phys.\ Lett.\ B {\bf 150} (1985), 301.

\bibitem{Matsuzaki:2015sya} 
  S.~Matsuzaki and K.~Yamawaki,
  arXiv:1508.07688 [hep-ph], to be published in JHEP.

    
\bibitem{Matsuzaki:2012mk} 
  S.~Matsuzaki and K.~Yamawaki,
  Phys.\ Lett.\ B {\bf 719}, 378 (2013);
  Phys.\ Rev.\ D {\bf 86}, 035025 (2012);
  Phys.\ Rev.\ D {\bf 85}, 095020 (2012).

\bibitem{Matsuzaki:2012xx} 
  S.~Matsuzaki and K.~Yamawaki,
  Phys.\ Rev.\ D {\bf 86}, 115004 (2012).
  
\bibitem{Sakata:1956hs}
  S.~Sakata,
  Prog.\ Theor.\ Phys.\  {\bf 16}, 686 (1956).
  
\bibitem{GellMann:1964nj}
  M.~Gell-Mann,
  Phys.\ Lett.\  {\bf 8}, 214 (1964);
  G.~Zweig,
  CERN-TH-401 (1964);
  PRINT-64-170 (1964).
    
\bibitem{Maki:1962mu}
  Z.~Maki, M.~Nakagawa and S.~Sakata,
  Prog.\ Theor.\ Phys.\  {\bf 28}, 870 (1962).
   
\bibitem{Kobayashi:1973fv}
  M.~Kobayashi and T.~Maskawa,
  Prog.\ Theor.\ Phys.\  {\bf 49}, 652 (1973).
 
\bibitem{Fukano:2015zua} 
  H.~S.~Fukano, S.~Matsuzaki, K.~Terashi and K.~Yamawaki,
  arXiv:1510.08184 [hep-ph].
 
 
\bibitem{Kurachi:2014qma} 
  M.~Kurachi, S.~Matsuzaki and K.~Yamawaki,
  Phys.\ Rev.\ D {\bf 90}, no. 5, 055028 (2014).
  
  
\bibitem{Bando:1984ej}
  M.~Bando, T.~Kugo, S.~Uehara, K.~Yamawaki and T.~Yanagida,
  Phys.\ Rev.\ Lett.\  {\bf 54} (1985), 1215;
  M.~Bando, T.~Kugo and K.~Yamawaki,
  Nucl.\ Phys.\  B {\bf 259} (1985), 493.
  
\bibitem{Bando:1987br}
  M.~Bando, T.~Kugo and K.~Yamawaki,
  Phys.\ Rept.\  {\bf 164} (1988), 217.

\bibitem{Harada:2003jx} 
  M.~Harada and K.~Yamawaki,
  Phys.\ Rept.\  {\bf 381}, 1 (2003).
   
\bibitem{Lee:2015qsa} 
  H.~K.~Lee, W.~G.~Paeng and M.~Rho,
  arXiv:1504.00908 [nucl-th];
  M.~Rho,
  arXiv:1507.05173 [hep-ph], in this Proceedings;  
  W.~G.~Paeng, T.~T.~S.~Kuo, H.~K.~Lee and M.~Rho,
  arXiv:1508.05210 [hep-ph].  
  
  
\bibitem{Matsuzaki:2013eva} 
  S.~Matsuzaki and K.~Yamawaki,
  Phys.\ Rev.\ Lett.\  {\bf 113}, no. 8, 082002 (2014)
  
\bibitem{Miransky:1988xi}
  V.~A.~Miransky, M.~Tanabashi and K.~Yamawaki,
  Phys.\ Lett.\  B {\bf 221}, 177 (1989);
  Mod.\ Phys.\ Lett.\  A {\bf 4}, 1043 (1989);
  
 \bibitem{Nambu:1989jt} 
  Y.~Nambu,
  Chicago preprint EFI 89-08 (Feb., 1989). Earlier related discussions are given in 
  H.~Terazawa, K.~Akama and Y.~Chikashige,
  Phys.\ Rev.\ D {\bf 15}, 480 (1977).
   
 \bibitem{Bardeen:1989ds} 
  W.~A.~Bardeen, C.~T.~Hill and M.~Lindner,
  Phys.\ Rev.\ D {\bf 41}, 1647 (1990).

     
\bibitem{Shuto:1989te}
  S.~Shuto, M.~Tanabashi and K.~Yamawaki,
 in {\it Proc. 1989 Workshop on Dynamical Symmetry Breaking}, 
      Dec. 21-23, 1989, Nagoya, eds. T. Muta and K. Yamawaki 
      (Nagoya Univ., Nagoya, 1990) 115-123;
  M.~S.~Carena and C.~E.~M.~Wagner,
  Phys.\ Lett.\  B {\bf 285}, 277 (1992);
  M.~Hashimoto,
  Phys.\ Lett.\  B {\bf 441}, 389 (1998).
  
\bibitem{Dobrescu:1997nm} 
  B.~A.~Dobrescu and C.~T.~Hill,
  Phys.\ Rev.\ Lett.\  {\bf 81}, 2634 (1998);
  R.~S.~Chivukula, B.~A.~Dobrescu, H.~Georgi and C.~T.~Hill,
  Phys.\ Rev.\ D {\bf 59}, 075003 (1999)

\bibitem{Fukano:2013aea} 
  H.~S.~Fukano, M.~Kurachi, S.~Matsuzaki and K.~Yamawaki,
  Phys.\ Rev.\ D {\bf 90}, no. 5, 055009 (2014);
  H.~C.~Cheng, B.~A.~Dobrescu and J.~Gu,
  JHEP {\bf 1408}, 095 (2014).

\bibitem{Fukano:2015ysa} 
  H.~S.~Fukano,
  arXiv:1507.08003 [hep-ph]. to appear in the SCGT15 Proceedings.
  
\bibitem{Kikukawa:1989fw} 
  Y.~Kikukawa and K.~Yamawaki,
  Phys.\ Lett.\ B {\bf 234}, 497 (1990).
  
\bibitem{Kondo:1991yk} 
  K.~i.~Kondo, S.~Shuto and K.~Yamawaki,
  Mod.\ Phys.\ Lett.\ A {\bf 6}, 3385 (1991);
  N.~V.~Krasnikov,
  Mod.\ Phys.\ Lett.\ A {\bf 8}, 797 (1993);
  K.~i.~Kondo, M.~Tanabashi and K.~Yamawaki,
  Prog.\ Theor.\ Phys.\  {\bf 89}, 1249 (1993);
  M.~Harada, Y.~Kikukawa, T.~Kugo and H.~Nakano,
  Prog.\ Theor.\ Phys.\  {\bf 92}, 1161 (1994).
  

\bibitem{ArkaniHamed:2001ca} 
  N.~Arkani-Hamed, A.~G.~Cohen and H.~Georgi,
  Phys.\ Rev.\ Lett.\  {\bf 86}, 4757 (2001);
  C.~T.~Hill, S.~Pokorski and J.~Wang,
  Phys.\ Rev.\ D {\bf 64}, 105005 (2001).


\bibitem{Son:2003et} 
  D.~T.~Son and M.~A.~Stephanov,
  Phys.\ Rev.\ D {\bf 69}, 065020 (2004);
  T.~Sakai and S.~Sugimoto,
  Prog.\ Theor.\ Phys.\  {\bf 113}, 843 (2005).
   
\bibitem{Matsuzaki:2015pda} 
  S.~Matsuzaki,
  arXiv:1510.04575 [hep-ph], to be published in the SCGT15 Proceedings.  
   
 \bibitem{Caswell:1974gg}
  W.~E.~Caswell,
  Phys. Rev. Lett. {\bf 33}, 244 (1974);
  T.~Banks and A.~Zaks,
  Nucl. Phys. B {\bf 196}, 189 (1982).
  
\bibitem{Appelquist:1996dq}
  T.~Appelquist, J.~Terning and L.~C.~Wijewardhana,
  Phys. Rev. Lett. {\bf 77}, 1214 (1996)
  T.~Appelquist, A.~Ratnaweera, J.~Terning and L.~C.~Wijewardhana,
  Phys. Rev. D {\bf 58}, 105017 (1998).
    
\bibitem{Yamawaki:2007zz} 
  K.~Yamawaki,
  Prog.\ Theor.\ Phys.\ Suppl.\  {\bf 167}, 127 (2007).


    
\bibitem{Hashimoto:2010nw} 
  M.~Hashimoto and K.~Yamawaki,
  Phys.\ Rev.\ D {\bf 83}, 015008 (2011).
  
\bibitem{Kaplan:2009kr} 
  D.~B.~Kaplan, J.~W.~Lee, D.~T.~Son and M.~A.~Stephanov,
  Phys.\ Rev.\ D {\bf 80}, 125005 (2009).

  
 
\bibitem{Dimopoulos:1979es} 
  S.~Dimopoulos and L.~Susskind,
  Nucl.\ Phys.\ B {\bf 155}, 237 (1979);
  E.~Eichten and K.~D.~Lane,
  Phys.\ Lett.\ B {\bf 90}, 125 (1980).
    
\bibitem{Yamawaki:1982tg}
  K.~Yamawaki and T.~Yokota,
  Phys.\ Lett.\  B {\bf 113} (1982), 293;
  Nucl.\ Phys.\  B {\bf 223} (1983), 144.
 
\bibitem{Gusynin:1987em} 
  V.~P.~Gusynin and V.~A.~Miransky,
  Phys.\ Lett.\ B {\bf 198}, 79 (1987)
  [Ukr.\ Fiz.\ Zh.\ (Russ.\ Ed.\ ) {\bf 33}, 485 (1988)].
  
\bibitem{Dimopoulos:1979sp} 
  S.~Dimopoulos,
  Nucl.\ Phys.\ B {\bf 168}, 69 (1980);
E.~Farhi and L.~Susskind,
  Phys.\ Rept.\  {\bf 74}, 277 (1981). 
  
\bibitem{Kurachi:2014xla} 
  M.~Kurachi, S.~Matsuzaki and K.~Yamawaki,
  Phys.\ Rev.\ D {\bf 90}, no. 9, 095013 (2014), and references cited therein.

\bibitem{Kurachi:2015bva} 
  M.~Kurachi, R.~Shrock and K.~Yamawaki,
  Phys.\ Rev.\ D {\bf 91}, no. 5, 055032 (2015).
 
\bibitem{Aoki:2015aqa} 
Y.~Aoki, T.~Aoyama, E.~ Bennett, M.~Kurachi, T.~Maskawa, K.~Miura, K.~-i.~Nagai, H.~Ohki and E.~Rinaldi,  A.~Shibata, K.~Yamawaki and T. Yamazaki  (the LatKMI Collaboration),  
  arXiv:1510.05863 [hep-lat];
  arXiv:1510.07373 [hep-lat]. To be published in SCGT15 Proceedings.
  
    
\bibitem{Aoki:2012eq} 
   Y.~Aoki, T.~Aoyama, M.~Kurachi, T.~Maskawa, K.~-i.~Nagai, H.~Ohki,  A.~Shibata, K.~Yamawaki and T. Yamazaki  (the LatKMI Collaboration),
  Phys.\ Rev.\ D {\bf 86}, 054506 (2012).
  
\bibitem{Aoki:2013xza} 
  Y.~Aoki, T.~Aoyama, M.~Kurachi, T.~Maskawa, K.~-i.~Nagai, H.~Ohki,  A.~Shibata, K.~Yamawaki and T. Yamazaki  (the LatKMI Collaboration),  
  Phys.\ Rev.\ D {\bf 87}, no. 9, 094511 (2013).
  
\bibitem{Appelquist:2014zsa} 
  T.~Appelquist {\it et al.}  [LSD Collaboration],
  Phys.\ Rev.\ D {\bf 90}, 114502 (2014);
%
  A.~Hasenfratz, D.~Schaich and A.~Veernala,
  arXiv:1410.5886 [hep-lat].
 
  

\bibitem{Aoki:2013zsa} 
Y.~Aoki, T.~Aoyama, M.~Kurachi, T.~Maskawa, K.~-i.~Nagai, H.~Ohki and E.~Rinaldi,  A.~Shibata, K.~Yamawaki and T. Yamazaki  (the LatKMI Collaboration),
  Phys.\ Rev.\ Lett.\  {\bf 111}, no. 16, 162001 (2013).

\bibitem{Aoki:2014oha} 
  Y.~Aoki, T.~Aoyama, M.~Kurachi, T.~Maskawa, K.~Miura, K.~-i.~Nagai, H.~Ohki and E.~Rinaldi,  A.~Shibata, K.~Yamawaki and T. Yamazaki  (the LatKMI Collaboration),
  Phys.\ Rev.\ D {\bf 89}, 111502(R) (2014);
PoS LATTICE 2013 (2013) 070.

\bibitem{Fodor:2014pqa} 
  Z.~Fodor, K.~Holland, J.~Kuti, D.~Nogradi and C.~H.~Wong,
  PoS LATTICE {\bf 2013}, 062 (2014);
  R.~Brower, A.~Hasenfratz, C.~Rebbi, E.~Weinberg and O.~Witzel,
  arXiv:1411.3243 [hep-lat].


\bibitem{Aad:2015owa} 
  G.~Aad {\it et al.}  [ATLAS Collaboration],
  arXiv:1506.00962 [hep-ex].
See also 
  V.~Khachatryan {\it et al.}  [CMS Collaboration],
  JHEP {\bf 1408}, 173 (2014). 
 
\bibitem{Fukano:2015hga} 
  H.~S.~Fukano, M.~Kurachi, S.~Matsuzaki, K.~Terashi and K.~Yamawaki,
  Phys.\ Lett.\ B {\bf 750}, 259 (2015).
 
 \bibitem{Fukano:2015uga} 
  H.~S.~Fukano, S.~Matsuzaki and K.~Yamawaki,
  arXiv:1507.03428 [hep-ph].
 

\end{thebibliography}


\end{document}